\documentclass[fleqn,usenatbib]{mnras}

\usepackage[T1]{fontenc}
\usepackage{ae,aecompl}
\pdfoutput=1 

\usepackage{graphicx}	
\usepackage{amsmath}	
\usepackage{amssymb}	

\title[Star--disc interaction in NGC~2264]{Evidence for disc regulation in the lowest-mass stars of the young stellar cluster NGC~2264}

\author[S. Orcajo et al.]{
Santiago Orcajo$^{1,2}$\thanks{E-mail: santi@fcaglp.unlp.edu.ar},
Lucas A. Cieza$^{3}$\thanks{E-mail: lucas.cieza@mail.udp.cl},
Roberto Gamen$^{1,2}$\thanks{E-mail: rgamen@fcaglp.unlp.edu.ar},
Dawn Peterson$^{4}$
\\
$^{1}$Facultad de Ciencias Astron\'omicas y Geof\'isicas, Universidad Nacional de La Plata, Paseo del Bosque S/N, 1900 La Plata, Argentina\\
$^{2}$Instituto de Astrof\'isica de La Plata, CCT La Plata, CONICET--UNLP, Argentina\\
$^{3}$N\'ucleo de Astronom\'ia, Facultad de Ingenier\'ia y Ciencias, Universidad Diego Portales, Av Ej\'ercito 441, Santiago, Chile\\
$^{4}$Space Science Institute, 4750 Walnut Street, Suite 205, Boulder, CO 80301, USA
}

\date{Accepted XXX. Received YYY; in original form ZZZ}

\pubyear{2015}

\begin{document}
\label{firstpage}
\pagerange{\pageref{firstpage}--\pageref{lastpage}}
\maketitle

\begin{abstract}

In the pre-main-sequence stage, star-disc interactions have been shown to remove stellar angular momentum and regulate the rotation periods of stars with M2 and earlier spectral types. Whether disc regulation also extends to stars with later spectral types still remains a matter of debate.  Here we present a star-disc interaction study in a sample of over 180 stars with spectral types M3 and later (corresponding to stellar masses $\leq 0.3 M_\odot$) in young stellar cluster NGC 2264. Combining rotation periods from the literature, new and literature spectral types, and newly presented deep \emph{Spitzer} observations, we show that stars with masses below 0.3  $M_\odot$ with discs also rotate slower than stars without a disc in the same mass regime. Our results demonstrate that disc-regulation still operates in these low-mass stars, although the efficiency of this process might be lower than in higher-mass objects. We confirm that stars with spectral types earlier and later than M2 have distinct period distributions and that stars with spectral types M5 and later rotate even faster M3 and M4-type stars.

%
\end{abstract}



\section{Introduction}\label{intro}

The angular momentum of a recently born star is one of the fundamental parameters such as the initial mass and chemical composition. Rotation influences the stellar internal structure, the energy transport, mass--loss, and the mixing of chemical elements.
Stars are supposed to inherit their own rotation from the natal molecular core, but this angular momentum is larger than the star could bear without breaking out itself \citep{1995ARA&A..33..199B}. 
Observations show that this does not happen, but speeds typically only reach up to a 10~\% of the break-up speed \citep[see e.g.][]{1986ApJ...309..275H}.
Thus, several processes have been proposed to slow down the stellar rotation, e.g. magnetic braking, discs, mass-loss, expanding envelopes, etc. \citep[see][and references therein]{2014prpl.conf..433B}.
In the case of low--mass (0.3-2.0 M$_{\odot})$ stars, it is believed that the major mechanism regulating the stellar rotation is the star-disc interaction.

The current paradigm on the evolution of low-mass pre-main sequence (PMS) stars indicates that the stars interact magnetically with their discs, removing angular momentum from the contracting stars and preventing them to reach break-up velocities as would it be expected from conservation of angular momentum alone \citep{2004AJ....127.1029R}. 
Protoplanetary discs have very varied lifetimes, ranging from less than a million years to 10 million years (Sicilia-Aguilar et al. 2006; Hernandez et al. 2007, Cieza et al. 2007).
%
The main prediction of this ``disc-braking" paradigm is that the stars that have lost their discs must rotate faster than those that still retain it. 

Early photometric measurements of rotation periods of PMS stars, combined with near-infrared (NIR) observations as a disc indicator (most commonly excess K-band emission), seemed to support the star-disc interaction scenario for angular momentum regulation. Rapidly rotating stars showed a lower mean NIR excess than did stars with longer rotation periods \citep{1993AJ....106..372E,2001ApJ...554L.197H,2002A&A...396..513H,2005A&A...430.1005L}. These early claims were challenged by later studies failing to confirm the correlation between rotation period and different disc and accretion indicators in a number of clusters
\citep{1999AJ....117.2941S,2001AJ....121.1676R,2004AJ....127.1029R,2004AJ....127.2228M,2010MNRAS.403..545L}. 

Most of the conflicting results can be traced back to the use of ambiguous disc indicators (e.g. $K$-band excesses) and the dependence of rotation periods on stellar mass. In particular, there is a sharp break in the behaviour of rotation period distributions for stars earlier and later than type M2.5 (corresponding to stellar masses of $\sim$0.2-0.3 M$_{\odot}$ 
at the age of a 1-3 Myr) \citep{2002A&A...396..513H,2010MNRAS.403..545L}, which requires to treat both groups of stars independently. Distinct rotation period distributions for different stellar mass regimes are also seen at older ages \citep[8-800 Myr;][]{2018AJ....155..196R}.
Also, because the contrast between the stellar photosphere and the thermal emission from the inner disc at NIR wavelengths is not as large as it is in the mid-IR, the near-IR observations miss 30~$\%$ of the discs that can be detected at longer wavelengths 
\citep{1998AJ....116.1816H}.
In addition, the contrast between the stellar photosphere and the emission from the inner disc in the NIR is greater for higher mass stars than it is for lower mass stars, since the stellar flux from the higher mass stars peaks at shorter wavelengths, causing a correlation between NIR excess and mass \citep{2005MNRAS.358..341L}. Solar-type (F8-K7) and early-M stars also inherently rotate more slowly than late-M objects. 
The combination of these two correlations can mimic the expected correlation caused by star-disc interaction, obscuring its signature if studied using NIR excess as a disc indicator.

Using mid-IR \emph{Spitzer} colors ([3.6]-[8.0]) as a robust disc indicator, \citet[][CB07, hereafter]{2007ApJ...671..605C} showed a clear correlation between rotation period and the presence of a disc in the Orion Nebula Cluster (ONC) and NGC 2264, for stars with spectral types M2 and earlier (M $\gtrsim$ 0.3 M$_{\odot}$). In particular, CB07 demonstrated that the bimodal distribution reported in the ONC in this mass regime by early observations \citep[e.g. ][]{2000ASPC..219..121H} is comprised of two distinct populations: stars without discs that have a period distribution peaked at $\sim$2 days and stars with discs with a period distribution peaked at $\sim$8 days.  CB07 also demonstrated that the  period distributions of stars with and without discs in NGC~2264 are significantly different 
in the sense that stars with a disc tend to rotate slower, as predicted by the disc regulation paradigm.
Using UV excess as an accretion disc indicator, \citet{2017A&A...599A..23V} confirms that discs influence the rotational evolution of stars in NGC 2264.

Whether disc-braking still operates in stars with spectral types M3 and later remains unclear. The \emph{Spitzer} data used by CB07 were not deep enough to robustly test disc-braking in the lowest-mass members from neither of their clusters. In particular,  the fact that the  8.0 $\mu$m data used does not reach the stellar photospheres of late M-type stars introduces a strong detection bias against the fast rotators that is difficult to interpret: the fast rotators could remain undetected either because they are the lowest mass stars in the sample or because they  are preferentially disc-less stars.
\citet{2006ApJ...646..297R} examined the relationship between \emph{Spitzer} mid-IR fluxes and rotation in stars between $0.1-3 ~M_\odot$ in ONC (using the same shallow \emph{Spitzer} adopted by CB07). They found that disc-hosting stars with estimated masses $< 0.25 ~M_\odot$ do rotate slower than the disc-less counterparts. 
However, in addition to the \emph{Spitzer} detection bias mentioned above, the study suffers from some ambiguity in the stellar mass estimate. Since they use of optical colours to identify stars more and less massive than 0.25$~M_\odot$, the low-mass sample can be contaminated by extincted higher-mass stars, which tend to rotate slower and have a higher \emph{Spitzer} detection rate.  
Subsequent disc regulation studies have produced mixed results.
\citet{2010A&A...515A..13R} find a disc-rotation connection in the  0.075 to 0.4~M$_{\odot}$ mass regime from a study of the $K$-band excesses of the periodic stars in the ONC, while \citet{2010ApJS..191..389C} find that stars in this mass regime with and without a disc have very similar period distributions in the $\sigma$-Ori cluster. 
\citet{2018ApJ...859..153S} also find  hints for disc regulation in the brown dwarf regime, although with a very small sample of only 25 objects. 

In this paper, we combine deep \emph{Spitzer} observations, with new spectral types of periodic stars in NGC 2264 to test disc-regulation paradigm on stars with masses below 0.3~M$_\odot$. 
Following CB07, we refer to these objects as ``low-mass" stars and to objects with spectral type M2 and earlier as ``high-mass'' stars. However, we emphasize that these are just \emph{relative terms}. Virtually all stars in NGC 2264 for which photometric periods have been obtained have solar or lower masses and can thus be considered ``low-mass'' stars. 
 
%
NGC~2264 is a young Galactic cluster in the Monoceros OB1 association in the Orion's Arm of our Galaxy.  It contains hundreds of young stars embedded in a large molecular cloud complex presenting diffuse H${\alpha}$ 
emission and differential interstellar extinction. 
The presence of Herbig-Haro objects and molecular flows confirm the active current star formation 
\citep{2005AJ....129..829D} and hence, its youth. 
\citet{2009AJ....138..963B} determined a distance of 913$\pm$40~pc and derived an age of $\sim$1.5~Ma. \citet{2008hsf1.book..966D} shows other estimates of ages to range between 1 and 5~Ma.
 NGC 2264 has about 400 objects with known rotation periods \citep{2004AJ....127.2228M, 2005A&A...430.1005L, 2013MNRAS.430.1433A, 2017A&A...599A..23V}. Of these 400 objects, an estimated 200 correspond to stars of spectral type M3 or later based on their optical colours.

In this work, we present a study of the rotational properties of a sample of 182 spectroscopically confirmed low-mass and 171 high-mass stars with and without discs.
The sample of periodic stars is discussed in Sec.~\ref{sample}.
The optical spectra used to characterize the periodic stars  and the \emph{Spitzer} data used for disc identification is discussed in Sec.~\ref{data}.
In Sec.~\ref{results}, we combine the \emph{Spitzer} data, the rotation periods, and the spectral types to search for correlations between
the presence of a disc and slow rotation, with a focus on stars with spectral types M3 and later. 
In Sec.~\ref{discussion}, we discuss our results and summarize our main conclusions.

\section{The periodic sample}\label{sample}



 Investigating the effect protoplanetary discs have on the rotation
periods of low-mass stars requires the combination of 3 sets of data:
rotation periods, a powerful disc indicator, and accurate spectral
types. Therefore, our analysis is restricted to stars with known rotations
periods from CB07, for which we have obtained deep \emph{Spitzer} data
for disk identification (see \ref{spitzer}) and optical spectra for
spectral typing (see \ref{spectra}). 
The rotation of solar and lower-mass stars can be determined by
photometric variations. These variations are produced by stellar spots
that modulate the optical flux as the star rotates. We adopt the
rotational periods from Makidon et al. (2004) and Lamm et al. (2005)
that were compiled by C07, and whenever possible, updated the periods
with more recent values from Affer et al. (2013) and Venuti et al. (2017).

%


%
%
%

\citet{2004AJ....127.2228M} observed NGC~2264 with the 0.76~m telescope at the McDonald Observatory, USA.
They obtained series of $I$-band CCD images, and detected 201 periodic variables in a sample of 4\,974 stars. They covered a 102 day baseline during the 1996--1997 fall--winter observing season. Based on their false alarm probability (FAP) levels, they divided their periods into two quality categories (1 and 2) and they recommended using the sample of 114 category-1 stars. 
 They considered all their periodic stars likely to be young stars and
therefore cluster members based on their position in the  I vs. V-I colour-magnitude diagram.

\citet{2005A&A...430.1005L} carried out a photometric monitoring program of a $34 \times 33$' field, with the Wide Field Imager (WFI) attached to the ESO/MPG 2.2~m telescope at La Silla, Chile. Their observations were obtained on 44 nights in the $I_{C}$ band between December 2000 and March 2001. They obtained typically 88 data points per object. Relative and absolute photometry was obtained on 10\,554 stars with magnitudes extending to $I_{C}$ $\sim$ 21 mag. 543 targets of 10\,554 were detected as periodic variables with a significance of 99\% or better.  Of these periodic variable stars, they classified 405 stars as PMS members based on their position on the $\mathrel {I_{{\rm C}}}$ vs. $\mathrel {(R_{{\rm C}}-I_{{\rm C}})}$ colour-magnitude-diagram and the
$\mathrel {(R_{{\rm C}}-{\rm H}\alpha )}$ vs. $\mathrel {(R_{{\rm
C}}-I_{{\rm C}})}$ colour-colour diagram.

\citet{2013MNRAS.430.1433A} obtained 189 rotational periods among NGC~2264 cluster members, based on the COnvection ROtation and planetary Transits (CoRoT) satellite. Observations were taken between March 7 to 31, 2008. 
They were able to reveal flux variations, with a precision down to 0.1~mmag. 
\citet{2017A&A...599A..23V} presented new rotation periods in NGC 2264 from additional CoRoT data taken between December 1, 2011 to January 9, 2012, showing a good agreement with periods from Affer (2013).


Following CB07, we combined the 114 quality-1 rotation periods from \citet{2004AJ....127.2228M} and the 405 rotation periods from \citet{2005A&A...430.1005L}.
For the 74 stars in common, we adopted the rotation periods from 
\citet{2005A&A...430.1005L}, resulting in a total of 445 periodic stars. 
Of these 445 stars,  163 have newer CoRoT rotation periods from \cite{2017A&A...599A..23V} and 19 from \cite{2013MNRAS.430.1433A}, which we consider more reliable and adopt over those of \citet{2004AJ....127.2228M} and \citet{2005A&A...430.1005L}.
However, we note that a large fraction ($>$50\%) of the periodic stars identified by ground-based observations are fainter than R $\sim$17 mag and were not monitored by CoRoT.
%
The rotation periods of the entire sample are listed in Table ~\ref{Tab445}


\section{New Observations \& Data}\label{data}

\subsection{\emph{Spitzer} data}\label{spitzer} 

\begin{figure}
	\includegraphics[width=1.2\columnwidth]{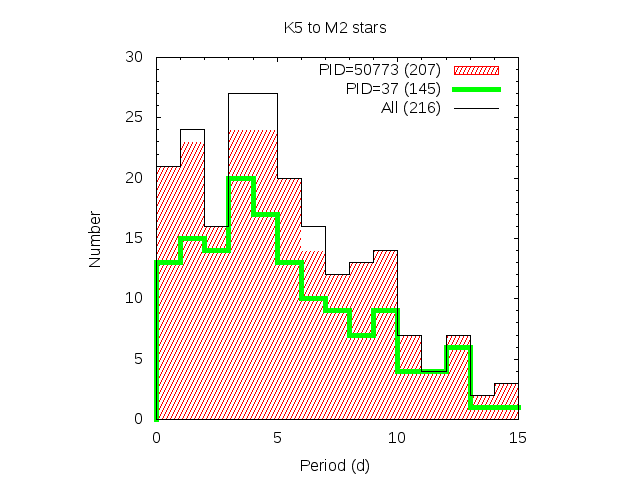}
        \includegraphics[width=1.2\columnwidth]{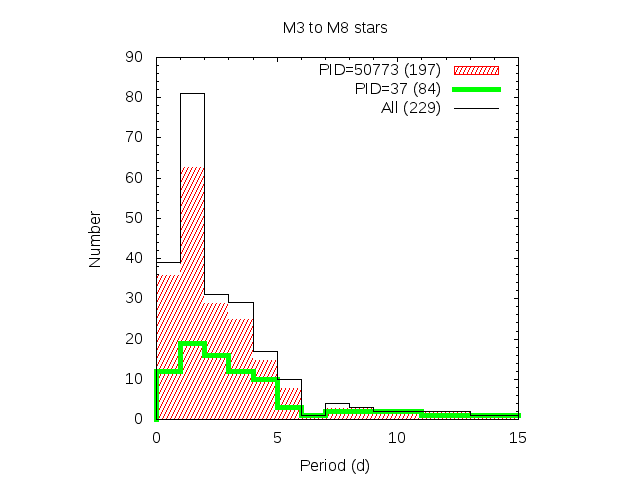}
	\caption{Period histograms for stars with and without 3.6 and 8.0 $\mu m$ \emph{Spitzer} photometry before and after including the Cycle-5 (PID = 50773) divided into K5 to M2 (top panel) and M3 to M8 (bottom panel) spectral types.  Following CB07, here K5 to M2 and M3 to M8 stars are discriminated by whether their colour index $R$--$I$ is lower  or greater than 1.3 magnitudes, respectively.
The three different lines represent all the periodic stars (thin black line), the stars detected by the \emph{Spitzer} program 50773 (filled red) and the \emph{Spitzer} program 37 (thick green line). 
 The \emph{Spitzer} data from program 37 detected only a small fraction of low-mass stars and was heavily biased against the fast rotators. 
This bias is mostly removed by the deeper data from \emph{Spitzer} program 50773. 
}
\label{PID}
\end{figure}

As mentioned in Sec.~\ref{intro}, the  IRAC (3.6 to 8.0 $\mu$m) data used by CB07 was not deep enough to test disc regulation in low-mass stars.  CB07 used archival data (PID = 37) obtained in the High Dynamic Range mode, which included 0.4~s observations and 10.4~s exposures at each dither position, for a total integration time of 43~s per pixel.  This relatively shallow data has a 5-$\sigma$ sensitivity of
0.1 mJy at 8.0 $\mu$m - and resulted in a low detection rate of low-mass stars (84 out 229 at 8.0~$\mu$m). Many of the low-mass stars that were not detected by \emph{Spitzer} are the fastest rotators, see Fig.~\ref{PID}. Moreover, many of these stars are the lowest mass stars in the sample, since they are the faintest. Because stars with circumstellar discs are brighter at 8.0 $\mu$m, they are easier to detect, causing a selection biased toward stars with discs. However, in order to understand the rotation period distribution of stars with and without discs in this mass range, the data must be close to complete down to stars with bare stellar photospheres for the entire mass range.

Here we present deep IRAC observations of a 32 x 32 arcminute field (PID = 50773) containing all the periodic stars analysed in CB07. The Astronomical Observation Requests (AORs) consist of a 7-row by  7-column map with a step size of 295 arcseconds. We achieved the required exposure time by dithering 30-second exposures 5 times. Therefore, each AOR amounts to a map with 150~s~pixel$^{-1}$ integrations, and we use 4 such AORs in total, for a total integration time of 600~s per pixel,  corresponding
to a 5-$\sigma$ sensitivity of 0.025 mJy at 8.0 $\mu$m . The data were taken on November 2nd, 2008, with AOR Keys  26896384, 26896640, 26896896, and 26897152. 

As in CB07, the data were processed using the pipeline developed as part of the \emph{Spitzer} Legacy Project, "From Molecular Cores to Planet-forming discs" \citep[c2d, ][]{2009ApJS..181..321E}. We merged the photometry tables from CB07 with the deeper data from the \emph{Spitzer} Cycle-5 program 50773.  In general, we adopt the data from the Cycle-5 program, except for the brightest objects ($>$ 6, 6, 46, and 24 mJy at 3.6, 4.5, 5.8, and 8.0 $\mu$m respectively) to avoid saturation. 
The merged \emph{Spitzer} photometry data is also listed in Table ~\ref{Tab445}. 
As seen in Fig.~\ref{PID}, the deeper \emph{Spitzer} data significantly improves the detection rate of the low-mass stars, especially the fast rotators. In particular, out of the 445 periodic stars, 439 are detected (with a signal to noise ratio $>$ 5) at 3.6 $\mu$m and 406 at 8.0 $\mu$m. In total, 404 periodic targets have [3.6] - [8.0] colours for disc identification.  






\subsection{Spectroscopic observations}
\label{spectra}

CB07 used optical colours to identify stars with M2 and earlier spectral types in their periodic sample.  In particular, they used an $R-I$ < 1.3 mag (corresponding to unextincted M2 stars and stars with earlier spectral types \citep{1995ApJS..101..117K} as the colour boundary. However, isolating low-mass stars in NGC~2264 is not possible from photometrical colour indexes due to the intrinsic differential extinction in the field (i.e.,
extincted early-type stars might have the $R-I$ colours of low-mas stars). 
Therefore, in order to construct an uncontaminated sample of low-mass stars, we employed low-resolution spectra, obtained from the Multi Mirror Telescope (MMT) and Gemini North (GN) telescopes, to determine spectral types and select the M-type stars for this work.

\subsubsection{MMT data}

We used the Hectospec spectrograph attached to the MMT to obtain the optical spectral of those members of the young stellar cluster NGC~2264 with known rotation periods and undetermined spectral types in order to construct a very clean sample of M3 and later type stars.
We employed the 270 groove/mm grating, which provides the complete optical spectrum 365--920~nm with resolution $R\sim1500$. Even though Hectospec has a total of 300 fibres, given the positioning constraints (due partly to a minimum distance between fibres of 20~arcsec and collision of fibre trajectories), three different fiber configurations were needed to observe $\sim$300 targets. For each fiber configuration, we take six sets of 15~min exposures. 
We also secured three sets of 5~min exposures shifted 5~arcsec off the program targets in order to optimize the sky subtraction. 
In total, we obtained 321 spectra in November 2009, under program 2008c-PA-08B-0209.  

The data were processed using a pipeline from the  Smithsonian Astrophysical Observatory Telescope Data Center, which  consists of a series of scripts based on standard IRAF\footnote{{\sc iraf} is distributed by the National Optical Astronomy Observatories, which are operated by the Association of Universities for Research in Astronomy, Inc., under cooperative agreement with the National Science Foundation.} routines (\textsc{dofibers} and \textsc{apextract} in particular) for CCD processing, aperture tracing, aperture extraction, wavelength calibration and sky substraction \citep[see][for details]{2005PASP..117.1411F,2005ASPC..347..228M}. 

\subsubsection{Gemini Data}
\label{gemini}

NGC~2264 was also observed with the Gemini Multi-Object Spectrograph (GMOS) in the multi-object mode from the Gemini North (GN) telescope (program ID GN-2009B-Q-76). 
We used the R400 grating, which delivers optical spectra between 550 and 1000~nm with a resolving power $R \approx 2000$.
Our GMOS targets have $R$-band magnitudes in the 18--20 range, and were integrated up a total of 20~min to reach typical Signal to Noise ratios ($S/N$) of 50. 
We were able to obtain 161 spectra in the 14 masks employed. Of these, 76 spectra correspond to periodic stars in CB07 and 84 stars are extra objects which are added in the Appendix Table~\ref{TableExtra}. 

Data were processed, in the standard way (CCD processing, aperture tracing, aperture extraction, wavelength calibration and sky substraction), using the {\sc GMOS-gemini} package within {\sc iraf}.

\section{Results}\label{results}

\subsection{Spectral characterization}

The spectroscopic data described in Sec.~\ref{spectra} were obtained to characterize the stellar sources. We classified the spectra by comparing the data with the standards published by Neill Reid in the web-page of the Space Telescope Science Institute (www.stsci.edu/$\sim$inr/ldwarf.html) and considering the criteria suggested by \cite{1991ApJS...77..417K}, \cite{1994AJ....108.1437H} and \cite{2015hsa8.conf..441A} for M-type stars, which are the focus of our study. 

The optical spectra of M dwarfs are dominated by molecular absorption bands mainly from titanium oxide (TiO) and vanadium oxide (VO). 
For all M-type stars the TiO band is prominent, but the VO band dominates from M5 to later types.

\begin{figure} \centering
    \includegraphics[angle=270,width=1.3\columnwidth]{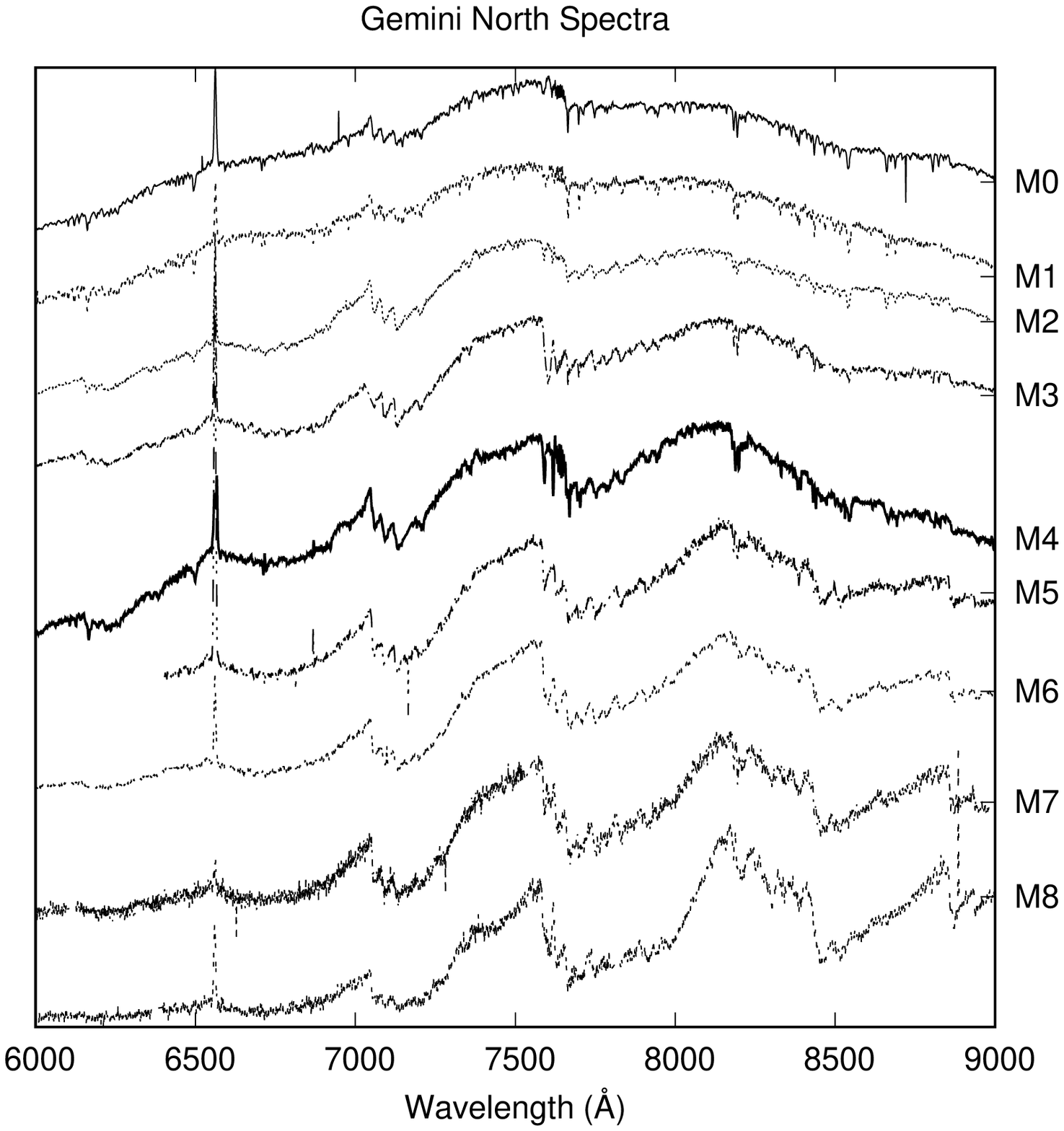}
    \includegraphics[angle=270,width=1.3\columnwidth]{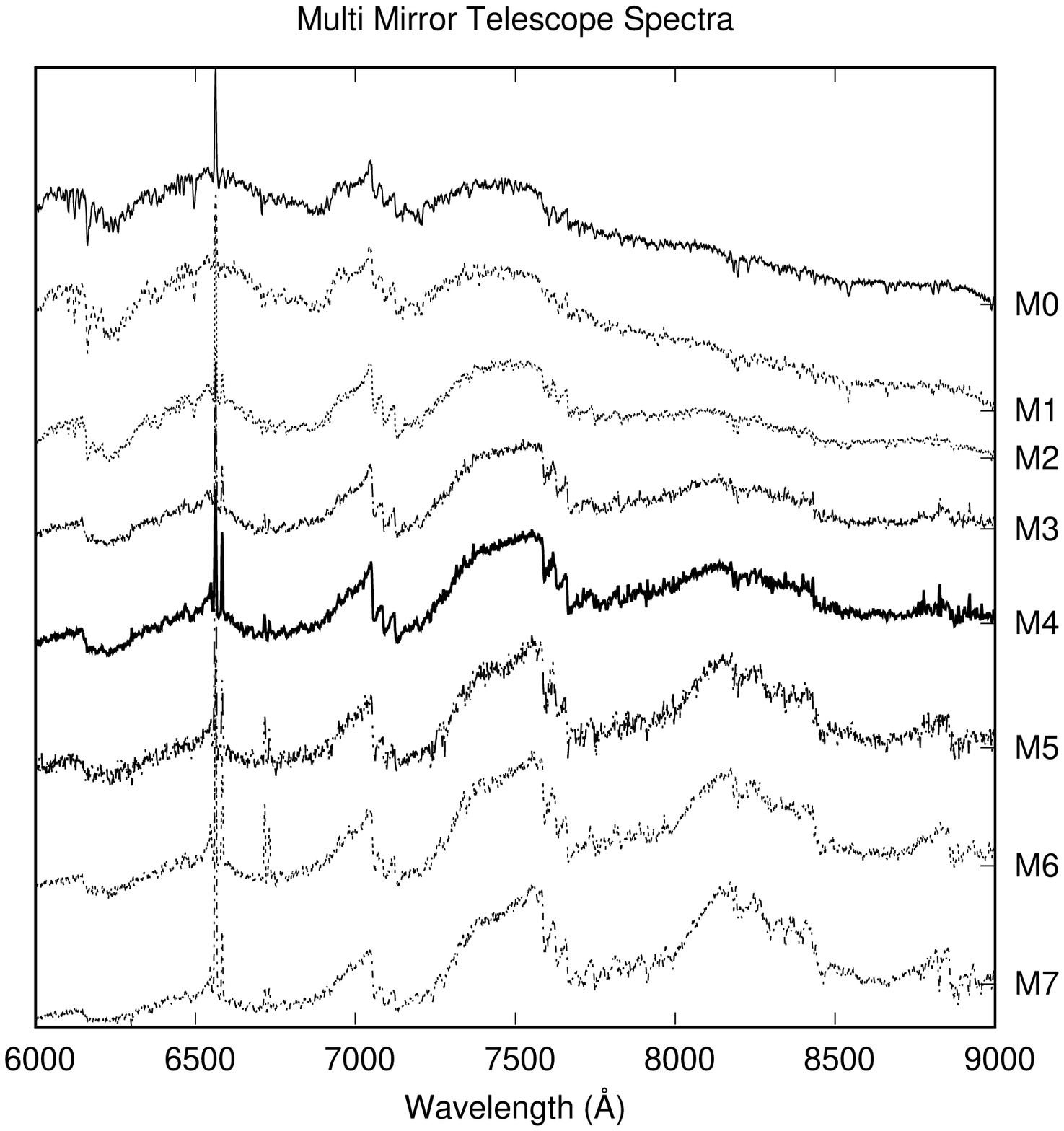}
    \caption{Our Gemini and MMT spectral sequences, from M0 to M8. The Gemini and MMT sequences have different spectral slopes due to differences in the instrumental responses and the pipeline processing.
}
	\label{espectros}
\end{figure}

We first constructed our own M-type spectral sequences from both Gemini and MMT data from a localized comparison to the characteristic features present in the published standards. The Gemini-GMOS and MMT-Hectospec spectral sequences for M-type stars are shown in Fig.~\ref{espectros}. Each sequence has its own particularities due to differences in the instruments and the details in the pipeline processing (e.g., sky subtraction and normalization). Also, with MMT we have not observed any objects with M8 spectral types.

For each of our spectra, we then performed a direct comparison against the corresponding spectral sequence (from Gemini or MMT). This was done combining an automatic script and visual inspection. We finally assigned a spectral type for each target based in the best match.

%
%
%

We determined the spectral types for 321 MMT targets and 161 Gemini targets. Since 29 objects were observed with both telescopes, we spectrally classified a total of 453 unique objects.
Of these objects, 365 have $\emph{Spitzer}$ and period data.  
%
Furthermore, 319 of these object are M-type stars.
%
The spectral class dataset was increased by 32 spectral types from the literature  provided by Rebull et al. (2002) and 30 by Skiff (2014), 14 and 20 of which correspond to M-type stars with \emph{Spitzer} and period data respectively.
Therefore, we used 353 M-type stars for this study, of which 171 are M0 to M2 spectral type and 182 are M3 to M8.
%
In general, we adopted the values by Rebull et al. (2002) over those of Skiff (2014) because the former presents homogeneous first-hand spectra, while the later presents data collected different literature  sources.
For M-type objects, we found 98 matching spectral types between Rebull et al. (2002) and ours, 83 of which agree within one sub-spectral type. Similarly, 67 stars have classification from both our own data and Skiff (2014). We find that 54 of them agree within 1 subtype of each other. We thus conclude that the three spectral classification are compatible within the uncertainties.
%
%

\begin{figure*}
    \includegraphics[width=1.0\columnwidth]{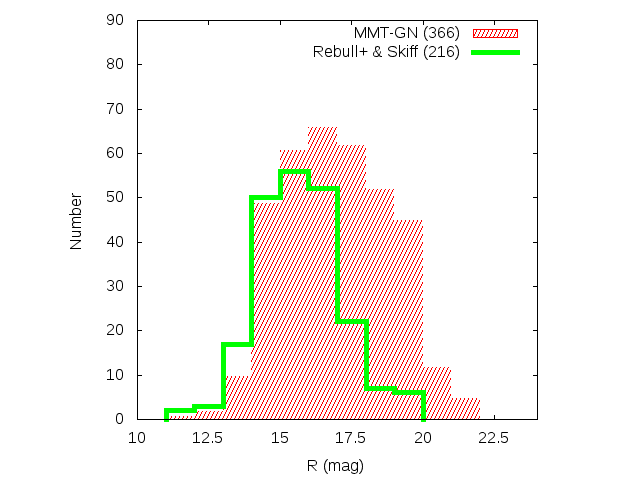}
    \includegraphics[width=1.0\columnwidth]{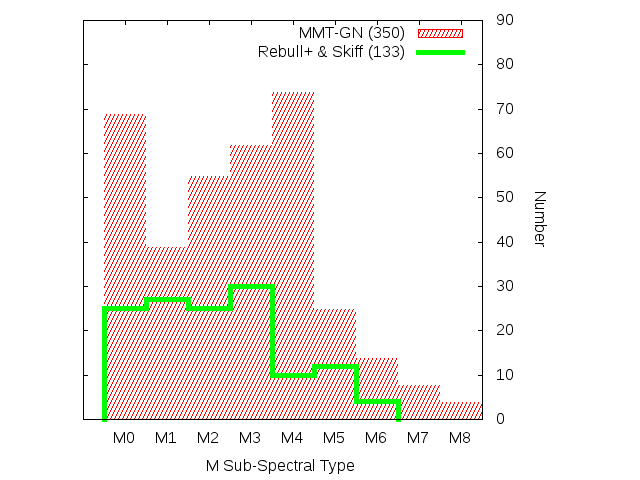}
    \caption{Distribution of $R$-band magnitudes (left) and spectral types (right) for the periodic stars before and after our spectral characterization with the MMT and Gemini observations.}
	\label{STLit}
\end{figure*}

%
In Fig.~\ref{STLit}, we show the distribution of $R$-band magnitudes for the periodic stars and indicate the sample of stars with known spectral types before and after our spectral characterization with the MMT and Gemini (top panel). We also show the distribution of spectral types, including values from the literature and our new results (bottom panel).
These figures show that we have significantly increased the sample of spectrally characterized low-mass stars, which are the key objects for this work.


Therefore, the sample used in this work comprises of 342 M-type stars, where 319 are spectral classifications made by us and 33 extracted from \citet{2014yCat....1.2023S}.
We consider our typical uncertainty to be one subtype.
These spectral types 
are listed in the column 15 of Table~\ref{Tab445}.
The spectral types of 85 objects that are not part of the periodic sample (typically bright objects that were assigned unused fibres or masks) are listed in the Table~\ref{TableExtra} in the Appendix.

\subsection{Disc identification}\label{disc_id}

Since very small amounts (M $<<$ M$_{\oplus}$) of warm dust (T $\sim$100--300 K) are needed to produce an optically thick mid-IR excess, most protoplanetary discs are optically thick at  \emph{Spitzer}-IRAC wavelengths (3.6 to 8.0 $\mu$m). 
Furthermore, the IRAC wavelengths are close to the Rayleigh-Jeans regime of the stellar photospheres and thus the [3.6]$-$[8.0] photospheric colour shows little dependence on spectral types. 
Therefore, \emph{Spitzer}-IRAC colours can be used as a robust disc indicator.  

\begin{figure*}
	\includegraphics[width=1.0\columnwidth]{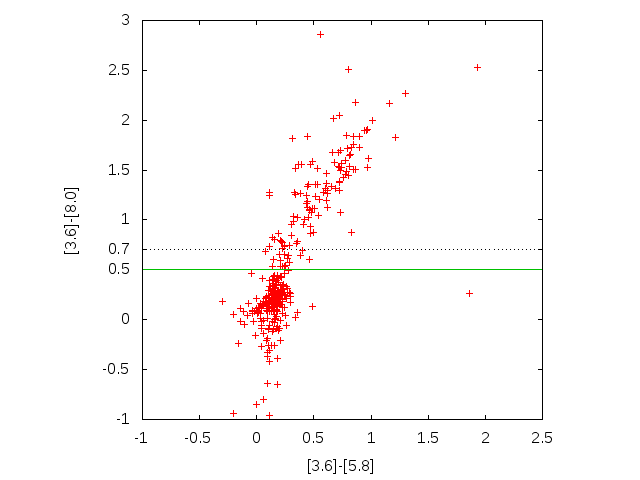}
	\includegraphics[width=1.0\columnwidth]{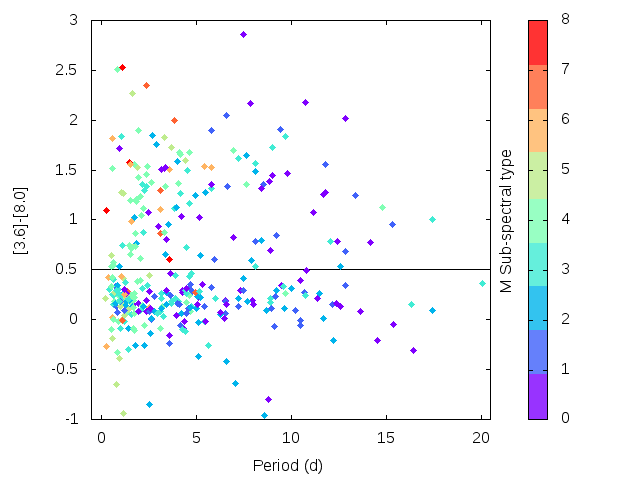}
	\caption{\textbf{Left panel:} The [3.6]$-$[8.0] vs [3.6]$-$[5.8] diagram of our M-type periodic stars. Note that most stars have colours $<$ 0.5 mag, consistent with bare stellar photospheres.
    \textbf{Right panel:} [3.6]$-$[8.0] vs period. Stars are colour-coded by spectral type. Later spectral type stars tend to have shorter periods than higher-mass objects.}
	\label{colorfig4}
\end{figure*}

Following CB07, we use \emph{Spitzer} colours to identify stars with discs. 
In Fig.~\ref{colorfig4} (left panel), we show the [3.6]$-$[8.0] vs [3.6]$-$[5.8] colours of our M-type periodic stars. We find that most stars have colours $<$ 0.5 mag, consistent with stellar photospheres. CB07 found that very few stars in their sample had  [3.6]$-$[8.0] colours between 0.5 and 0.7, and used  [3.6]$-$[8.0] $>$ 0.7 mag as a conservative disc identification criteria. We found 16 M-stars with [3.6]$-$[8.0] colours in the 0.5--0.7 mag range, and given the sharp paucity of objects at [3.6]$-$[8.0] = 0.5 mag in NGC~2264, we adopt this later boundary for disc identification purposes. 
 We note that the disc identification criterion is only weakly
affected by extinction as the median A$_V$ of 4.3 mag in NGC 2264
(Teixeira et al. 2012) corresponds to a reddening of just 0.08 mag in
the [3.6]-[8.0] colour (Chapman et al. 2009).

 In Fig.~\ref{colorfig4} (right panel), we show the [3.6]$-$[8.0] colours vs rotation period, colour-coded by spectral type. In the next section, we use these three parameters (\emph{Spitzer} colour, period and spectral type) to investigate disc regulation in the low-mass regime. 

Thanks to our deeper \emph{Spitzer} data and new spectra, we were able to obtain \emph{Spitzer}-IRAC colours for 404 periodic sources with known spectral types. In particular, we establish that 77 of the 182 low-mass stars (M3 and later types) have a disc, while the other 105 have colours consistent with bare stellar photospheres.
%

%


\subsection{Testing the disc regulation in the low-mass regime}


\begin{figure}
    \includegraphics[width=1.0\columnwidth]{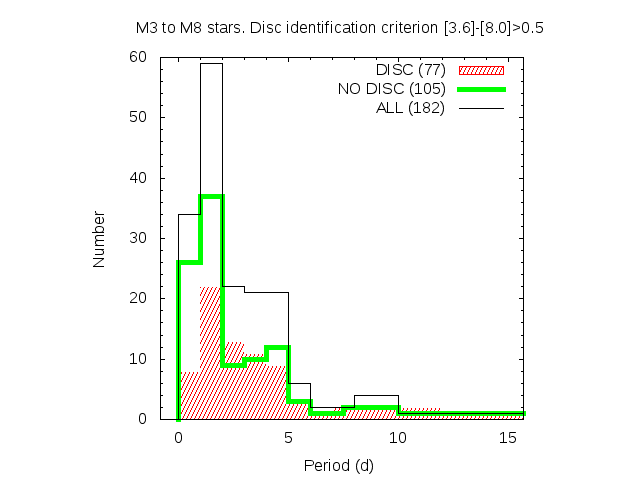}    
    \caption{Results for low-mass stars in NGC~2264. The  period distributions for low-mass stars with and without a disc using [3.6]-[8.0] $>$ 0.5 as disc identification criterion. The three different lines represent all the stars (thin black line), stars with a disc (filled red line) and stars without disc (thick green line). The period distribution of disc-less low-mass stars are shifted toward shorter periods. These distributions are significantly different   ($P=1.8E-4$, Kolmogorov-Smirnov two sample test). This result suggest that stars without a disc are free to spin up faster than stars with discs.  
}
	\label{lowdisc}
\end{figure}    

\begin{figure*}
    \includegraphics[width=1.0\columnwidth]{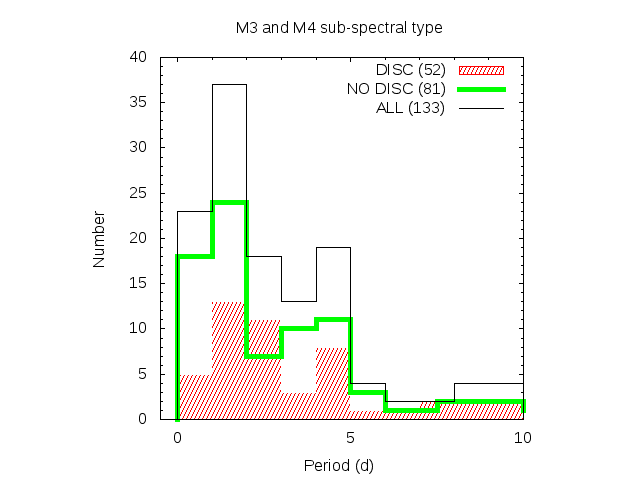}   
    \includegraphics[width=1.0\columnwidth]{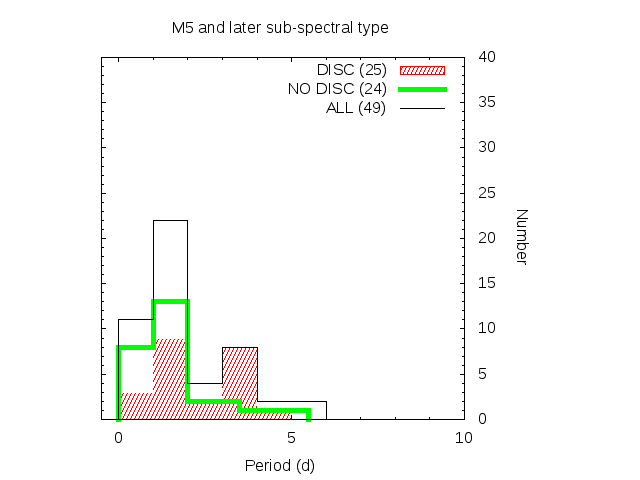}
    \includegraphics[width=1.0\columnwidth]{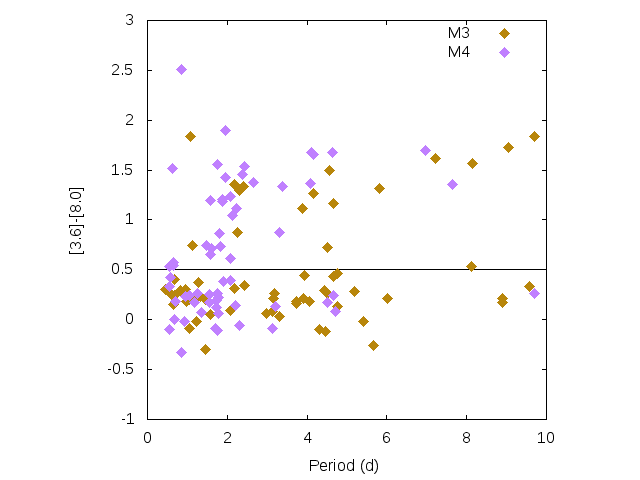}   
    \includegraphics[width=1.0\columnwidth]{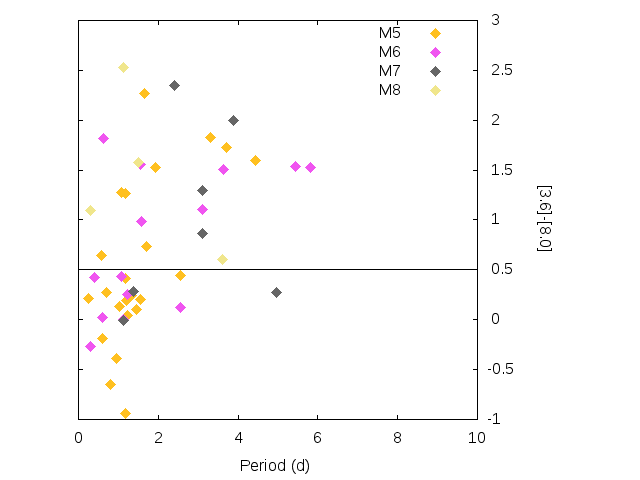}
    \caption{
\textbf{Top-left Panel}:  Period histograms for M3 and M4 stars with (52 stars) and without (81 stars) discs. The three different lines represent all the stars (thin black line), stars with a disc (filled red line) and stars without disc (thick green line). 
The period distribution for M3 and M4 stars concentrates at short periods ($P<5$~d) but extends to 10 days.\textbf{Top-right Panel}: The same period histograms for M5 and later-type stars with (25) and without (24) disc. The period distribution for M5 \& late stars is restricted to $P<5$~d. These stars are faint, making it difficult to detect them at \emph{Spitzer} wavelengths. 
\textbf{Lower Panel:}  [3.6]-[8.0] vs period for both mass regimes. Stars are colour-coded by spectral
type.  
The figures show the disc-period and disc-stellar mass dependencies extend to the entire sample.}

	\label{M3458}
\end{figure*}




The spectral characterization and disc identification discussed above allow us to test the disc regulation paradigm in the low-mass regime (spectral types M3 and later) for the first time. In total, we have obtain mid-IR and spectral types for 182 low-mass periodic stars.
In Fig.~\ref{lowdisc}, we show the period distribution of the low-mass stars with and without a disc using the disc identification criterion discussed in Section~\ref{disc_id} ([3.6]-[8.0] $>$ 0.5). 
We find that the distribution of periods for stars without a disc is displaced toward shorter periods with respect to that of the stars with a disc. 
In particular, discless stars have  mean and median periods of 2.72 and 1.56 days, while objects with disc have mean and median periods of 3.85 and 2.38 days.
In order to quantify the significance of these differences, we perform a Kolmog\'orov-Smirnov (KS) two-sample test and find that the maximum deviation in the normalized cumulative distributions,  $d$, is  0.31, while there is only a 1.8E-4 probability that both distributions are drawn from the same parent distribution ($p$ value). 
%
Using the slightly more conservative disc identification criterion adopted by CB07 ([3.6]-[8.0] $>$ 0.7), we find similar statistics for the KS test  ($d$ = 0.35 and $p$ = 2.4E-5). 

In Fig.~\ref{M3458} (top panels), we plot the low-mass stars divided into two subsamples: objects with spectral types M3 and M4 (left panel), and objects with spectral types M5 and later (right panel). 
 We also plot [3.6]-[8.0] as a function of period (bottom panels) for both mass regimes. 
Despite the smaller sample sizes,  the results for the K-S tests are still significant even for the latest spectral types considered  (M5 to M8 stars, $d$ = 0.59 and $p$ = 1.6E-4). 

\begin{figure}
    \includegraphics[width=1.2\columnwidth]{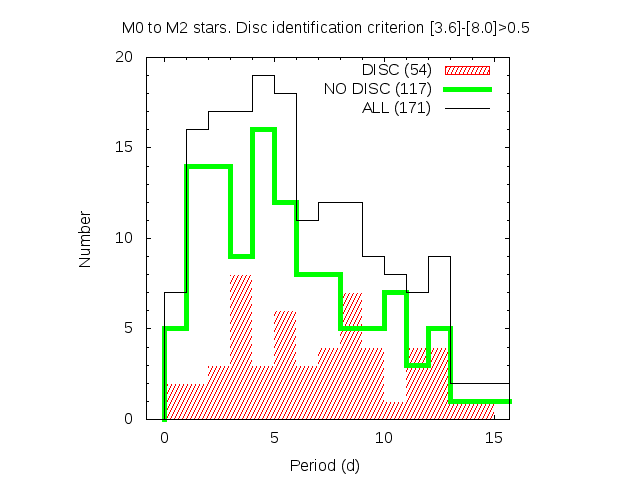}
    \caption{The same period distributions as in Fig~5, but but for spectral types M2 and earlier stars in NGC~2264.}
	\label{highmass}
\end{figure}

For completeness, we also perform the analysis of ``high-mass" stars using the spectrally-identified stars with spectral types M2 and earlier to confirm the results presented by CB07. In Fig.~\ref{highmass}, we show the distribution periods of ``high-mass" stars with and without discs and confirm that stars with discs have a  relatively flat distribution between 1 and 15 days, while stars without discs tend to have periods shorter than $\sim$5~d.  The K-S test provides the following statistics $d$ = 0.22 and $p$ = 0.04. 

In Table 2, we present the statistics from all the K-S tests for the different subsamples using both [3.6]-[8.0] $>$ 0.5  and [3.6]-[8.0] $>$ 0.7 as disc identification criteria. In general, we find that stars that have already lost their discs always rotate significantly faster than stars  that still retain a protoplanetary disc, as expected from the disc regulation paradigm. \\

\begin{table*}\tiny
\begin{center}
\begin{tabular}{cccccccccccccccc}
 \hline
 \hline
RA & DEC & Rc & Ic & Flux\_3.6 & err\_3.6 & Flux\_4.5 & err\_4.5 & Flux\_5.8 & err\_5.8 & Flux\_8.0 & err\_8.0 & Per & Ref1 & ST & Ref2 \\
 (J2000) &   (J2000)     &      (mag)	&  (mag)	&   (mJy)	&   (mJy)	&   (mJy)	&   (mJy)  	&   (mJy)	&   (mJy)	&   (mJy)	&   (mJy)       &  (d)	&         	&         &        \\
\hline
100.16500 &	 9.29761 &	 16.13 & 14.91 & 3.62e+00 &	 4.38e-02 &	 2.33e+00 &	 3.02e-02 &	 1.60e+00 &	 2.27e-02 &	 9.17e-01 &	 2.66e-02 &	 9.66 	& Mak 	& M2 	& 1 \\
100.05875 &	 9.34131 &	 13.14 & 12.64 & 9.88e+00 &	 1.11e-01 &	 6.23e+00 &	 6.93e-02 &	 3.96e+00 &	 5.07e-02 &	 2.46e+00 &	 3.78e-02 &	 1.292 	& Ven 	& Early & 1 \\
100.04958 &	 9.35281 &	 15.48 & 14.40 & 4.00e+00 &	 1.16e-01 &	 2.37e+00 &	 8.27e-02 &	 1.59e+00 &	 4.66e-02 &	 8.99e-01 &	 3.13e-02 &	 5.423 	& Ven 	& M3 	& 1 \\
100.06625 &	 9.35931 &	 16.14 & 14.89 & 3.66e+00 &	 5.37e-02 &	 2.59e+00 &	 3.11e-02 &	 1.65e+00 &	 3.25e-02 &	 9.63e-01 &	 2.84e-02 &	 0.643 	& Ven 	& M3 	& 1 \\
100.25700 &	 9.38436 &	 19.13 & 17.10 & 8.11e-01 &	 1.84e-02 &	 5.31e-01 &	 9.72e-03 &	 3.46e-01 &	 4.59e-02 &	 1.44e-01 &	 2.53e-02 &	 0.29 	& Lam 	& M6 	& 1 \\
\hline
\end{tabular}
\caption{NGC~2264 stars with periods from the literature and \emph{Spitzer} Cycle-5 data. Notes: - Ref1. are as follows:  Ven indicates periods and optical photometry taken from Venuti et al. (2017), Aff indicates periods and optical photometry taken from Affer et al. (2013),  Lam indicates periods and optical photometry taken from Lamm et al. (2005), while  Mak indicates periods and optical photometry taken from Makidon et al. (2004). Ref2  are as follows: 1 indicates our spectral types,  2 indicates spectral types taken from Rebull et al. (2002) and 3 indicates spectral types taken from Skiff (2009-2016). Objects that are not M-types were not classified and are simply labelled as ``Early''. The printed edition contains only a sample to illustrate its content. The complete version of this table is in the electronic edition of the Journal.  The data shown correspond to Spitzer program 50773, except for the
brightest objects (> 6, 6, 46, and 24 mJy at 3.6, 4.5, 5.8, and 8.0 m
respectively), for which we adopt the fluxes from Spitzer program
50773 to avoid saturation.
}
\label{Tab445}
\end{center}
\end{table*}

\begin{table}
\begin{center}
\begin{tabular}{lccc}
 \hline
 \hline
  Samples, disc criterion$^{a}$ &Sample sizes  & $d$  & $p$ \\
                                 &(disc/no-disc)&        & \\
 \hline
M2 and earlier,0.5 &	54, 117   &   0.22    &  2.4E-2 \\
M3 and later,  0.5 &	77, 105	  &   0.31    &  1.8E-4 \\
M3 and M4,     0.5 &	52, 81	  &   0.27    &  1.2E-2 \\
M5 and later,  0.5 &    25, 24    &   0.59    &  1.6E-4 \\  
\hline 
M2 and earlier, 0.7  &	46, 125   &   0.22     &  6.3E-2 \\
M3 and later,   0.7  &	69, 113	  &   0.35     &  2.4E-5 \\
M3 and M4,      0.7  &	46, 87	  &   0.36     &  1.1E-3 \\
M5 and later,   0.7  &  23, 26    &   0.59     &  1.9E-4 \\
\hline
\end{tabular}
\caption{KS statistics for different samples. $a$: the disc criterion corresponds to the [3.6]-[8.0] color boundary used for disc identification.}
\label{Table:KS}
\end{center}
\end{table}



\section{Discussion and Conclusions}\label{discussion}

\subsection{Discussion}

The results presented in the previous section represent the clearest evidence to date that low-mass stars with discs rotate slower than stars that have already lost their discs. However, the overall period distribution in this regime is characterized by fast rotators, with a peak at $\sim$2 days and few stars with rotation periods longer than $\sim$5 days. Several stars even approach the breakup limit (where the centrifugal force equals gravity), which for PMS stars corresponds to rotation periods of $\sim$0.1 to 0.2 days ~\citep{2014prpl.conf..433B}.

The short rotation of low-mass stars is, at least in part, due to their smaller radii, $R$. As discussed by ~\citet{2010ApJS..191..389C}, since the specific angular momentum $j$ is related to the rotation  period ($P$) as $j$ $\propto$ R$^{2}$/P,  smaller stars with a given specific angular momentum will necessarily rotate significantly faster than their larger counter parts. The shorter periods in low-mass stars could in principle also be due to a lower efficiency in the disc-braking mechanism. In fact, the sharp break in the behaviour of the rotation period distributions for stars earlier and later than M2 suggests a change in the disc-braking mechanism at that boundary rather than smooth dependence of rotation period on stellar mass. 

Investigating the efficiency of disc-braking as a function of stellar mass requires numerical simulations as those presented by ~\citet{2017A&A...600A.116V}. They  compare Monte Carlo simulations to the observed period distributions of stars of different masses and ages, with and without discs, in order to constrain the initial period distributions, disc lifetimes, and disc locking efficiencies. Based on these comparisons,  they conclude that disc breaking is less important in low-mass stars and brown dwarfs
than in solar-mass stars. They speculate that the reduced efficiency of disc braking  might be due to changes in the topology of the magnetic fields in the low-mass regime. 
The sample of periodic low-mass stars with robust disc indicators presented herein will allow  to perform more detailed comparisons to numerical models and set quantitative constraints on the efficiency of disc-braking as a function of stellar mass. 





\subsection{Summary and Conclusions}

In order to test the disc-regulation paradigm in low-mass stars (M3 and later types) of the young open cluster NGC~2264, we have combined rotation periods from the literature with deep \emph{Spitzer} data (used as a disc indicator) and new optical spectroscopic data from Gemini and the MMT. 
We have obtained spectral types for 453 stars, 230 of which are periodic low-mass stars.
From our \emph{Spitzer} observations, we have obtained [3.6]-[8.0] colours for 404 objects, 
182 of which are periodic low-mass stars.  
From the analysis of these data, we can reach the following conclusions:


\begin{itemize}

\item We confirm that stars with spectral types earlier and  later than M2 have distinct period distributions.  Stars with M2  and earlier spectral types have a wide distribution of 
rotation periods, peaking at $\sim$5 days and extending to $\sim$15 days, while stars with M3 and later types have a period distribution peaking at $<$ 2 days and very few stars with periods longer than 5 days. 

\item We find that the dependence of rotation period on (sub)stellar mass extends to the lowest-mass objects in our sample: stars with spectral types M5 and later rotate even faster M3 and M4-type stars (mean periods are 1.9 and 3.62 days, respectively). 

\item We find, for the first time, clear evidence for disc-braking in the low-mass regime: stars with discs rotate slower than stars without a disc. When considering all stars with M3 and later types, this trend is seen with very high significance ($p$ $\sim$ 10$^{-5}$ according to the K-S test). The same trend is still  seen when sub-samples are considered (M3 and M4-types or M5 to M8-type objects), although with an slightly lower level of significance due to the smaller sample sizes. We also confirm the importance of disc-braking in stars with M2 and earlier spectral types. This suggest that disc-braking operates at all (sub)stellar masses. 

\item With a large sample of spectroscopically characterized periodic stars with deep mid-IR observations, NGC 2264 represents an ideal laboratory for quantitative studies of disc regulation as a function of stellar mass. 

\end{itemize}




\section*{Acknowledgements}
We thank the anonymous referee whose comments and suggestions have
helped to improve the paper significantly. L.A.C. acknowledges support
from CONICYT-FONDECYT grant 1171246.




\bibliographystyle{mnras}
\bibliography{discos} 

\appendix
\section{Appendix}\label{appendix}

\subsection{Extra table}
\begin{table*}[Extra target with only spectral type.]
\begin{center}
\begin{tabular}{ccccccc}
 \hline
 \hline
 RA      &  DEC    & ST     & H$\alpha$     & EW     & Err   & Ca II          \\
(J2000)  & (J2000) & 	    & (Em=1/Ab=0) & (\AA)    & (\AA) & (Em=1/Ab=0)   \\
\hline
100.2842 & 9.5693  & M4	    & 1           & -16.19 & 0.63    &   0           \\
100.2654 & 9.47223 & M2	    & 1           & -2.20  & 0.04    &   0           \\
100.2877 & 9.48535 & M7	    & 1           & -89.57 & 6.63    &   1           \\
100.4697 & 9.49029 & Early	& 0           &        &         &   0           \\
100.5022 & 9.4927  & Early	& 0           &        &         &   0           \\
100.3821 & 9.50017 & Early	& 0           &        &         &   0           \\
\hline
\end{tabular}
\caption{Additional spectroscopic targets not corresponding to the periodic sample and not necessarily members of NGC 2264. They are typically bright objects 
that were assigned unused fibres or masks. The printed edition contains only a sample to illustrate its content.
Objects that are not M-types were not classified and are simply labelled as ``Early''.  
}
\label{TableExtra}
\end{center}
\end{table*}

\bsp	
\label{lastpage}
\end{document}